\documentclass[prl,twocolumn,superscriptaddress,showpacs]{revtex4}
\usepackage[english]{babel}
\usepackage{amsmath}
\usepackage{amssymb}
\usepackage{graphicx}
\begin{document}
\newcommand{\lf}{\left}
\newcommand{\rg}{\right}
\newcommand{\be}{\begin{equation}}
\newcommand{\ee}{\end{equation}}
\newcommand{\bea}{\begin{eqnarray}}
\newcommand{\eea}{\end{eqnarray}}
\newcommand{\nn}{\nonumber}
\newcommand{\ba}{\begin{array}}
\newcommand{\ea}{\end{array}}
\newcommand{\op}[1]{#1}
\newcommand{\co}[1]{\overline{#1}}
\newcommand{\adj}[1]{#1^{\dagger}}
\newcommand{\al}[1]{\left(#1\right)}
\newcommand{\bra}[1]{\langle#1|}
\newcommand{\ket}[1]{|#1\rangle}
\newcommand{\dint}{\int\!\!\!\!\int}
\newcommand{\arccot}{\mathrm{arccot}}
\newcommand{\Tr}{\mathrm{Tr}}
\newcommand{\diag}[1]{\mathrm{diag}\al{#1}}
\newcommand{\re}{\mathrm{Re}}
\newcommand{\im}{\mathrm{Im}}
\renewcommand{\a}{\alpha}
\renewcommand{\b}{\beta}
\newcommand{\g}{\gamma}
\newcommand{\G}{\Gamma}
\newcommand{\D}{\Delta}
\renewcommand{\d}{\delta}
\newcommand{\ve}{\varepsilon}
\newcommand{\z}{\zeta}
\newcommand{\h}{\eta}
\renewcommand{\k}{\kappa}
\renewcommand{\l}{\lambda}
\newcommand{\n}{\nu}
\renewcommand{\r}{\rho}
\newcommand{\vf}{\varphi}
\newcommand{\F}{\Phi}
\renewcommand{\o}{\omega}
\renewcommand{\O}{\Omega}
\newcommand{\s}{\sigma}
\title{Resonant all-electric spin pumping with spin-orbit coupling}

\begin{abstract}
All-electric devices for the generation and filtering of spin currents are of crucial importance for spintronics experiments and applications.
Here we consider a quantum dot 
between two metallic leads 
in the presence of spin-orbit coupling,  
and we analyze in the frame of a scattering matrix approach  
the conditions for generating spin currents in an adiabatically driven two-terminal device. 
We then focus on a dot with two resonant orbitals and show by specific examples 
that both spin filtering and pure spin current generation can be achieved.
Finally, we  discuss the effect of the Coulomb interaction.

\end{abstract}

\author{Valentina Brosco}
\affiliation{Dipartimento di Fisica, Universit\`a ``La Sapienza'', P.le 
A. Moro 2, 00185 Roma, Italy}
\affiliation{Institut f\"ur Theoretische Festk\"orperphysik, Karlsruhe Institute of Technology, 
76128 Karlsruhe, Germany}

\author{Markus Jerger}
\affiliation{Physikalisches Institut, Karlsruhe Institute of Technology, 
76128 Karlsruhe, Germany}

\author{Pablo San-Jose}
\affiliation{Institut f\"ur Theoretische Festk\"orperphysik, Karlsruhe Institute of Technology, 
76128 Karlsruhe, Germany}
\affiliation{Instituto de Ciencia de Materiales de Madrid, CSIC, Sor Juana In\'es de la Cruz 3, E28049 Madrid, Spain}

\author{Gergely Zarand}
\affiliation{Theoretical Physics Department, Institute of Physics, 
Budapest University of Technology and Economy, Budapest, H-1521, 
Hungary} 
\affiliation{DFG Center for Functional Nanostructures (CFN), Karlsruhe Institute of Technology, 
76128 Karlsruhe, Germany}

\author{Alexander Shnirman}
\affiliation{Institut f\"ur Theorie der Kondensierten Materie, 
Karlsruhe Institute of Technology, 76128 Karlsruhe, Germany}
\affiliation{DFG Center for Functional Nanostructures (CFN), Karlsruhe Institute of Technology, 
76128 Karlsruhe, Germany}

\author{Gerd Sch\"on}
\affiliation{Institut f\"ur Theoretische Festk\"orperphysik, Karlsruhe Institute of Technology, 
76128 Karlsruhe, Germany}
\affiliation{DFG Center for Functional Nanostructures (CFN), Karlsruhe Institute of Technology, 
76128 Karlsruhe, Germany}

\maketitle 

{\it Introduction.} Many theoretical and experimental 
efforts have been devoted to spintronics, i.e.,~the design and control of spin-based 
electronic devices \cite{wolf01}, and 
as a result it became possible to inject and filter spin 
polarized  currents \cite{jedema}, to detect spin accumulation \cite{potok02,lou07, frolov}, and to produce ferromagnetic spin valves \cite{ralph07}.
While an all-electrical 
control of spin currents would be clearly advantageous, 
so far most theoretical  and experimental designs involve ferromagnetic leads or require the 
application of external magnetic fields. 

 Adiabatic pumping of charge in a cyclically 
modulated potential was first proposed by Thouless \cite{thouless} and later studied  
in a variety of mesoscopic devices~\cite{buttiker94,brouwer98,avron00,makhlin,moskalets03}. 
More recently, pumping of spins in nanostructures has been proposed as well, 
again with most mechanisms relying on external 
magnetic or exchange fields~\cite{mucciolo, governale,brouwer03,aono,citro06,schiller08},
or the presence of ferromagnetic leads~\cite{splett08}, and indeed in one experiment spin pumping in a magnetic field has been observed~\cite{watson}. 
On the other hand, as first discussed in Refs.~\cite{governale,brouwer03}, it is also possible to pump spin 
through quantum wires in the absence of 
external magnetic fields provided spin and orbit are coupled. 

 In the present 
paper we  show that, in the presence of spin-orbit (SO) coupling, 
resonances associated with avoided level crossings of a quantum dot can be exploited
to {\em pump } spin in a controlled  way purely by cycling electrical gates.
We shall focus on quantum dots with parameters such that the level spacing 
of the dot exceeds the typical width of the levels, 
$\delta\epsilon > \Gamma$. In this parameter regime individual states act as 
resonances, with position and coupling to the 
external leads which can be tuned by external gates \cite{zumbuhl}.  
 {\em Resonant spin pumping} 
emerges, when two of the levels lie close to the Fermi 
energy so that the SO coupling mixes them. We will show that 
 in the vicinity of such resonant avoided level crossings 
 the quantum dot can be used as an all-electric spin battery, 
and pumping cycles with transmitted spin of order $\hbar/2$ per cycle can be constructed.
 We also show that it is possible to choose the 
cycle parameters such that no charge is transfered through the quantum dot. 

{\em Scattering formalism.} We  consider an elastic 
scatterer coupled to  two  quasi-1D
leads. In the left lead, far from the 
scattering region we can define  
longitudinal charge and spin current operators, $J^0_L$ and $J^i_L$,  as 
\be
\label{Jc} J_{L}^\mu (z,t)=-\frac{i}{2m}\int \!dx\,dy\, 
\lf[\psi^{\dag}_L \s^\mu \big(\partial_z\psi_L\big)-{\rm h.c.}\rg],
\ee
%
%
%
where  $\s^0$ is the unit matrix, 
while $\s^\mu$  with $\mu=1,2,3$ denote the usual Pauli matrices. 
The spinor field
  $\psi_L(x,y,z,t)$  destroys an electron in the 
left lead and  $\partial_z$ denotes the partial derivative with 
respect to the coordinate along the wire. 
  
  At low temperatures, we can use
the approach of Brower \cite{brouwer98,brouwer03},
to express the spin 
($\vec S_{L}$) and charge ($Q_{L}$) 
pumped into the left lead within an 
adiabatic cycle  as  
\begin{subequations}
\begin{align}
\label{brouwerC}
 & Q_L=-\frac{e}{2\pi}
 \int_0^T \im \left[ \Tr \left\{ (\Lambda_L\otimes\sigma_{0})\, \frac{d {\cal S}}{dt}\adj{{\cal S}} \right\}\right]dt, \\
 \label{brouwerS}
 & \vec S_L=-\frac{\hbar}{2\pi} \int_0^T \im \left[
 \Tr \left\{(\Lambda_L\otimes\vec \sigma)\, \frac{d {\cal S}}{dt}\adj{{\cal S}}\right\}\right]dt. 
\end{align}
\end{subequations}
Here ${\cal S}={\cal S}(E_F,t)$ is  the instantaneous scattering  
matrix at the Fermi energy, and $\Lambda_L$ stands for the projector onto the left channel.

Following a strategy similar to Avron \emph{et al.}~\cite{avron00},
we decompose the  scattering matrix to 
identify  physical processes  which  contribute to spin and charge pumping. 
The presence of time-reversal symmetry implies that, in an appropriate basis,   
the scattering matrix is self-dual~\cite{beennaker97},  
i.e., ${\cal S}=\sigma_{y}{\cal S}^{T} \sigma_{y}$,
and, for a quantum dot connectung two one-mode leads, ${\cal S}$ can be decomposed as
\begin{equation}
    \label{Sdecompose}
    {\cal 
S}  =U^{0} U \, T \, \adj{U} U^{0}
    \end{equation}
where the matrices $U$, $U^0$ and $T$ are defined as follows:
 \bea U^0 & =& \left(
                  \begin{array}{cc}
                    e^{i \phi_{L}} & 0 \\
                    0 & e^{i \phi_{R}} \\
                  \end{array}
                \right)\otimes \sigma_{0}, \qquad\,\,
 U = \left(
                  \begin{array}{cc}
                   {\cal U}_L & 0 \\
                    0 & {\cal U}_R \\
                  \end{array}
                \right),\nn\\[0.3cm]
 T & =& \left(
        \begin{array}{cc}
            -\sqrt{1-T_0} & \sqrt{T_0}\\
            \sqrt{T_0} & \sqrt{1-T_0}\\
        \end{array}
    \right)\otimes \sigma_{0}. \label{UU0T}
\eea 
where $T_0$ denotes the transmission coefficient of the dot and ${\cal U}_{L,\,R}$ are  two-dimensional unitary matrices.

Substituting Eqs. (\ref{Sdecompose}-\ref{UU0T}) into 
Eqs. (\ref{brouwerC}-\ref{brouwerS}) we obtain
\begin{subequations}
\begin{align}
\label{Ic} & Q_{L}  =\frac{e}{2\pi} \int_0^T\!\!\left[
(1-T_0) 
\left(\dot{\phi}_{R}-\dot{\phi}_{L}\right)\right] dt,\\
\label{Is} & \vec S_{L}= \frac{i\hbar}{2\pi}\int_0^T \!T_0\,\, \Tr 
\left\{  \lf(\adj{{\cal U}}_{L}\,\vec \sigma\,{\cal U}_L\rg)\left( 
\adj{{\cal U}}_{L} \dot{{\cal U}_{L}} -\adj{{\cal U}}_{R} \dot{{\cal 
U}_{R}} \right)\right\} dt.
\end{align}
\end{subequations} 
In Eq.~\eqref{Ic}
 we dropped a term 
$(e/2\pi)\int  (\dot{\phi}_{R} + \dot{\phi}_{L})dt$, 
representing the charge accumulated on the scatterer,
which vanishes over a complete cycle. 
%
The first term in  \eqref{Ic} is finite even 
for 
$T_0=0$. This limit corresponds to peristaltic pumping, 
whereby a charge is first moved from the left lead to the dot, 
while the right contact is kept closed,  and subsequently from the dot to 
the right lead with the left contact kept closed.   

Interestingly, the pumped spin 
given by Eq.~(\ref{Is}) is proportional to $T_0$.
%
Therefore peristaltic spin pumping is not possible in 
the presence of time reversal symmetry. Also, 
while over a full cycle the charge  is conserved, $Q_L+Q_R=0$, the spin in general is not conserved: in 
particular if $[{\cal U}_L,{\cal U}_R]\neq0$ we have $\vec 
S_{R}+\vec S_{L}\neq 0$. This is not surprising, since 
spin-orbit coupling  -- ultimately responsible for the spin pumping -- 
allows for the transfer of angular momentum to the underlying lattice. 
We remark that the pumped spin   in Eq.~(\ref{Is})  transforms as a vector under spin rotations in the left lead 
while it is invariant under spin rotation in the right lead, as it was  also clear from the definition in Eq.  (\ref{Jc}).

%
\begin{figure}[b]
\centering
\includegraphics[width=7cm]{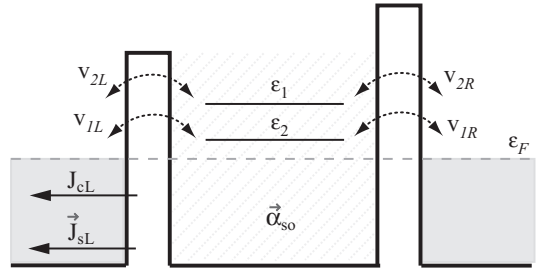}
\caption{Orbital level structure of the quantum dot coupled to the leads. Due to the spin orbit field, $\vec \a_{so}$, pumping gives rise to charge and a spin current.}\label{fig:dot}
\end{figure}
{\em Non-interacting quantum dot.}  To calculate explicitly the  
charge and spin transferred through a dot in the vicinity of 
an avoided level crossing,  let us now introduce a non-interacting
model to describe a dot with two orbital levels, 
$\ket{n\s}$ 
($n=1,2$, $\s=\pm$) as shown in Fig.\ref{fig:dot}. 
In the presence of (SO) coupling the isolated dot 
is  described by the  Hamiltonian $H^d = \sum d^\dagger_{n\s}
{\cal H}^d_{n\s,n'\s'}d_{n'\s'}$, with the operators 
$d^\dagger_{n\s}$ creating an electron in state $\ket{n\s}$
and the $4\times4$ matrix ${\cal H}^d$ given by
\be 
\label{hd}
   {\cal H}^{d}= \left(
  \begin{array}{cc}
    \ve_1\,\s^0 & 
    -i\vec{\a}\cdot\vec{\sigma} 
    \\
    i\vec{\a}\cdot\vec{\sigma} & 
    \ve_2\,\s^0
  \end{array}
  \right)\;.
\ee
Here $\ve_n$ indicate the energies of the dot orbital states 
measured from the Fermi energy and  the real vector 
$\vec \a$ is the  SO field. Choosing the spin 
quantization axis parallel  
 to $\vec \a$, we can express the full
Hamiltonian as $H = \sum_{\s}H_{\s}$ with 
\bea
H_{\s} = \sum_{n}\;\ve_n d^{\dag}_{n\s}d_{n\s}\;+
i\;\a \;s_\s \lf(d_{1\s}d^{\dag}_{2\s}-h.c.\rg) \label{h}
\\
\phantom{nn}+  \sum_{ n, \xi,\l} \lf(v^\l_n 
   \;c_{\xi\s\l}^{\dag}d_{n\s}+h.c.\rg)+\sum_{\xi,\l}\xi \;c^{\dag}_{\xi\s\l}c_{\xi\s\l}
\; .
\nn
\eea
with  $s_\s=\pm 1$ for spin parallel/antiparallel to $\vec \a$ and $\a=|\vec \a|$.
Here $c^{\dag}_{\xi\s\l}$ creates a conduction electron of 
energy $\xi$ and spin $\s$ in lead $\l$. In the following, we shall 
assume that while the levels $\ve_n$ and the tunneling amplitudes $v^\l_{n}$ 
can  be tuned via gate voltages, $\vec \a$ remains 
approximately constant over a pumping cycle. 
Choosing $\vec \a$ as a quantization axis we can write the scattering matrix in  spin-diagonal form, 
${\cal S}=e^{i(\phi_{L}+\phi_{R})}{\cal S}_{\uparrow}\oplus {\cal S}_{\downarrow}$, with  $S_{\s}$ parametrized using $\phi=\phi_{L}-\phi_{R}$
as 
\be
   {\cal S}_{\sigma}  =
\\
\left(
    \begin{array}{cc}
       {-e}^{i\phi}\sqrt{1-T_0} & 
        e^{is_\s\,\theta}\sqrt{T_0}
        \\
        e^{-is_\s\,\theta}\sqrt{T_0} &
         e^{-i\phi}\sqrt{1-T_0}
        \end{array}
    \right)\;.
\ee
Obviously, the phase $\phi$  is related to charge-pumping, while the 
phase $\theta$ determines the amount of spin pumped into the leads. 
If we only modify two external parameters, $r_1$ and $r_2$,
during a cycle,  we can use Stokes theorem to
recast the pumped charge and spin as
\bea
    \label{Qc}
    Q_L & = &
    -\frac{e}{2\pi}\,\dint {\rm d}^2{r}\; B^{r_1r_2}_c
\;,
\\ 
\label{Sz}
    \vec S_L& = &
  \hat \a \; \frac{\hbar}{4\pi}\,\dint {\rm d}^2{r}\;B^{r_1r_2}_s \;,
 \eea
where the charge and spin Brouwer's fields are defined as 
$B^{r_1r_2}_c = {\partial}_{r_1} T_{0} {\partial}_{r_2}\phi- {\partial}_{r_2} T_{0} {\partial}_{r_1}\phi $ and $B^{r_1r_2}_s = {\partial}_{r_1} T_{0} {\partial}_{r_2}\theta- {\partial}_{r_2} T_{0} {\partial}_{r_1}\theta$, respectively.

In the rest of the paper we shall study  specific pumping cycles 
with two pumping parameters,  either
the tunnel couplings or the positions of the dot levels, appearing in the Hamiltonian $H$. 
To use Eqs.~\eqref{Qc} and \eqref{Sz}, we  
need to compute the transmission $T_0$ and the angles $\phi$
and $\theta$ in terms of these quantities. This can be done most easily  by relating 
${\cal S}$ to the Green's function $G^{d}$ of the dot~\cite{ng88}, 
\be
     {\cal S}^{\l\l'}_{\s} = 
    \delta_{\l\l'} - 2\pi i\sqrt{\r_\l\r_{\l'}}\sum_{n,n'}v^\l_n 
    {v^{\l'}_{n'}}^*\;G_{n,n'\s}^{d}(\ve_F)\;, \nn
\ee
\be
\left[G^{d}(\omega)\right]_{nn',\s}^{-1}  =
\omega \;\delta_{nn'}-\left({\cal H}_{\s}^{d}\right)_{nn'} + i\pi \sum_{\l} \rho_\l 
{v^\l_n}^* v^\l_{n'}\;.
\nn
\ee
where  $\rho_\l$ is the 
density of states in lead $\l$ at the  Fermi energy.
From these equations  we can  obtain  $T_0$, $\phi$, and $\theta$  
in terms of the bare parameters of the dot. Introducing $w=v_{1L} v_{2R} -v_{2L} v_{1R}$
we can express for $\phi$ and $\theta$ as follows
%
\bea \label{theta}
    \tan(\theta) &=&
    \frac{
      \a\;  w}
    {\ve_1 \;v_{2L} v_{2R} +\ve_2\; v_{1L} v_{1R}}\;,
 \\
    \tan(\phi) &=&
    \pi\rho 
        \frac{\ve_1(v_{2L}^{2}-v_{2R}^{2}) + \ve_2(v_{1L}^{2}-v_{1R}^{2})}
        {\ve_1\ve_2-\a^{2}+\pi^{2}\rho^{2}w^{2}}\label{phi}\;. 
\eea
Clearly, both levels are 
involved in spin pumping:  if one of the levels 
is decoupled from the leads then  $w=0$, and
only charge pumping is possible. Coupling  the first and 
second level to the two leads with equal amplitudes
also leads to a vanishing of the spin current. 

In the following we analyze two kinds  of cycles, ``orbital energy cycles'' and 
 ``tunnel coupling cycles'', 
where only the orbital energies of the levels are varied or also the tunnel coupling to the leads. 
Spin filtering and pure spin pumping can be achieved in both cycles.   

{\em Orbital energy cycles.}  
%
%
\begin{figure}[b]
\begin{minipage}[b]{4cm}
\centering
\includegraphics[width=4cm]{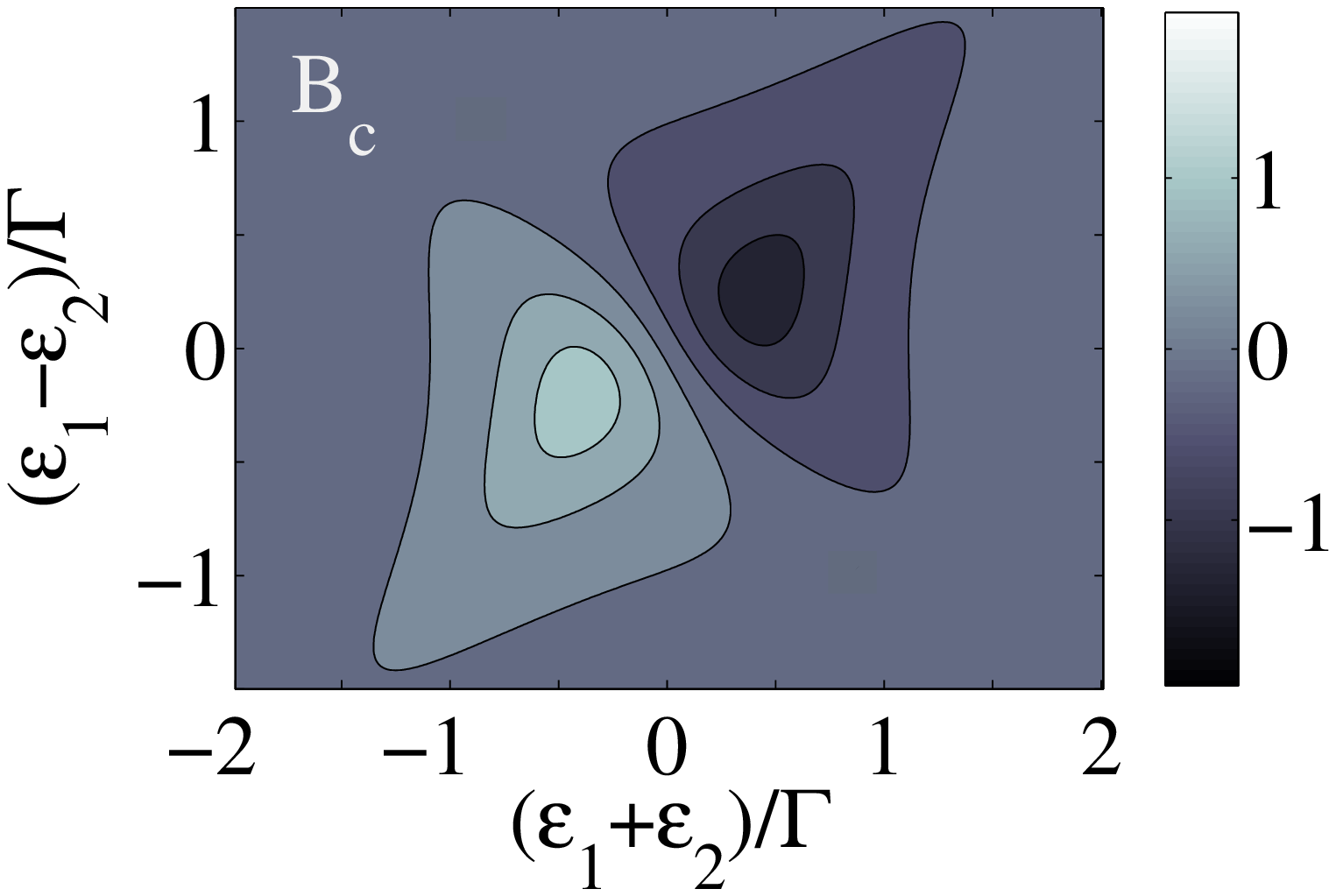}
\end{minipage}
\begin{minipage}[b]{4cm}
\centering
\includegraphics[width=4cm]{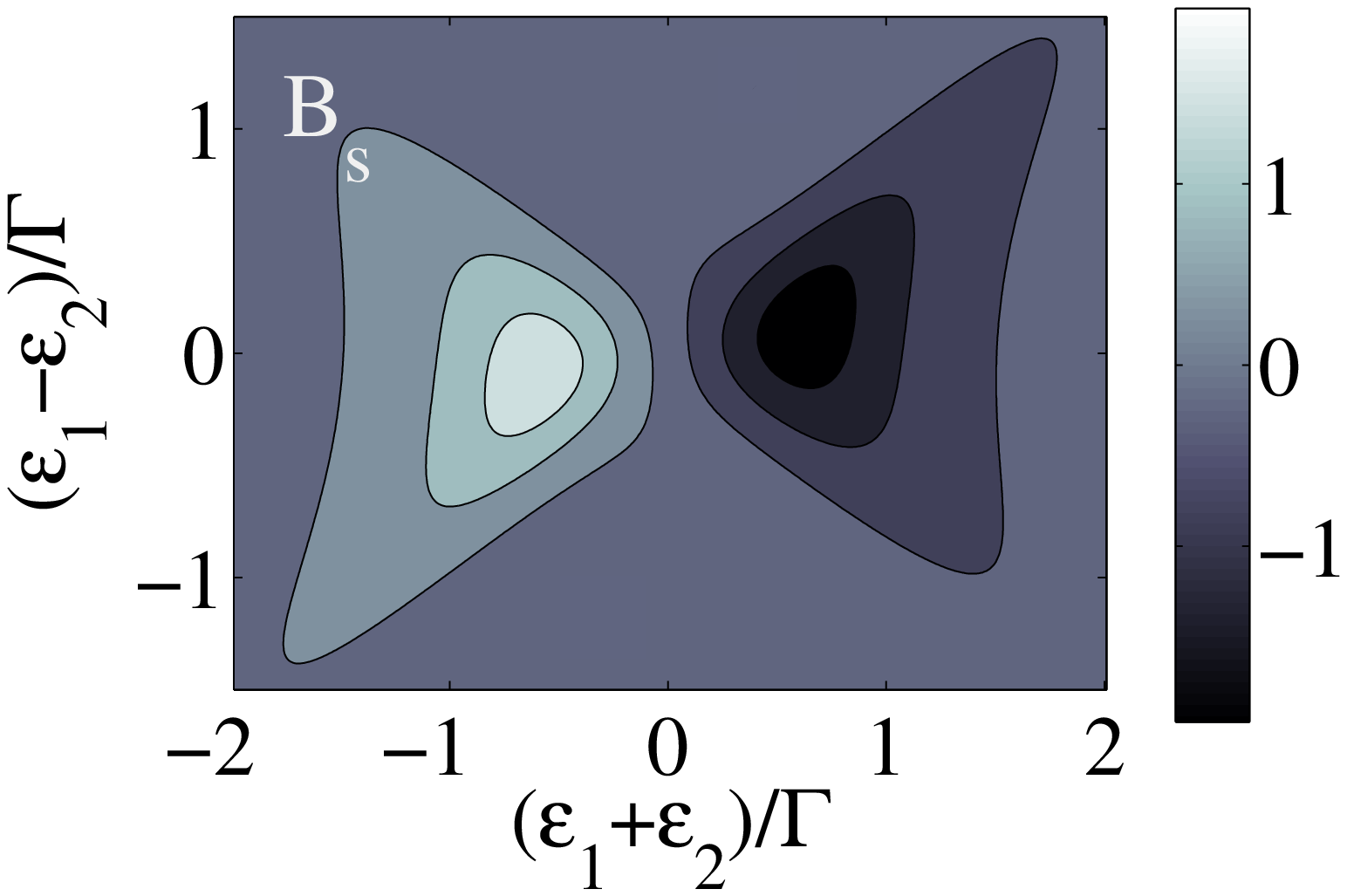}
\end{minipage}
\begin{minipage}[b]{4cm}
\centering
\includegraphics[width=4cm]{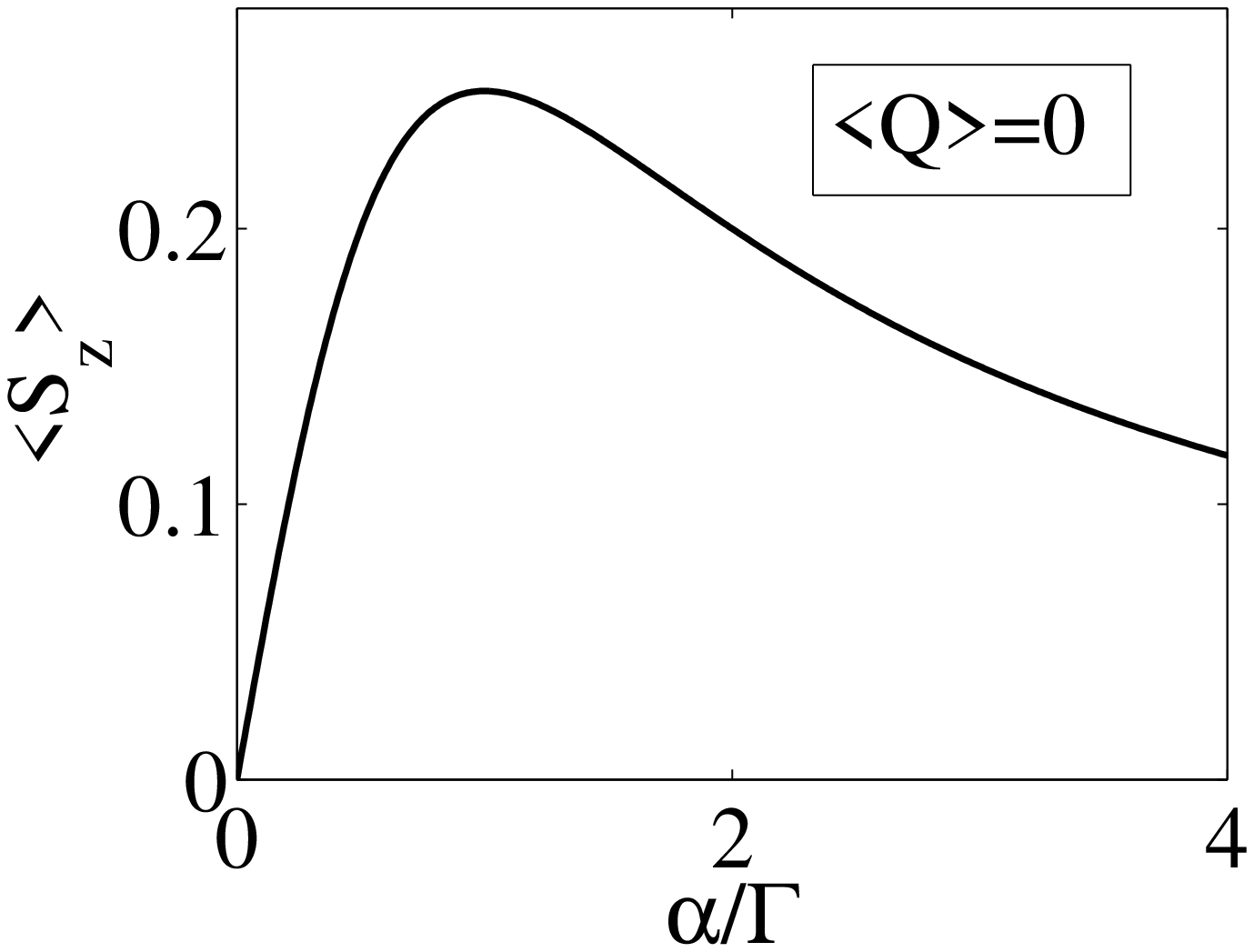}
\end{minipage}
 \hspace{-3mm} 
\begin{minipage}[b]{4cm}
\centering
\includegraphics[width=4cm]{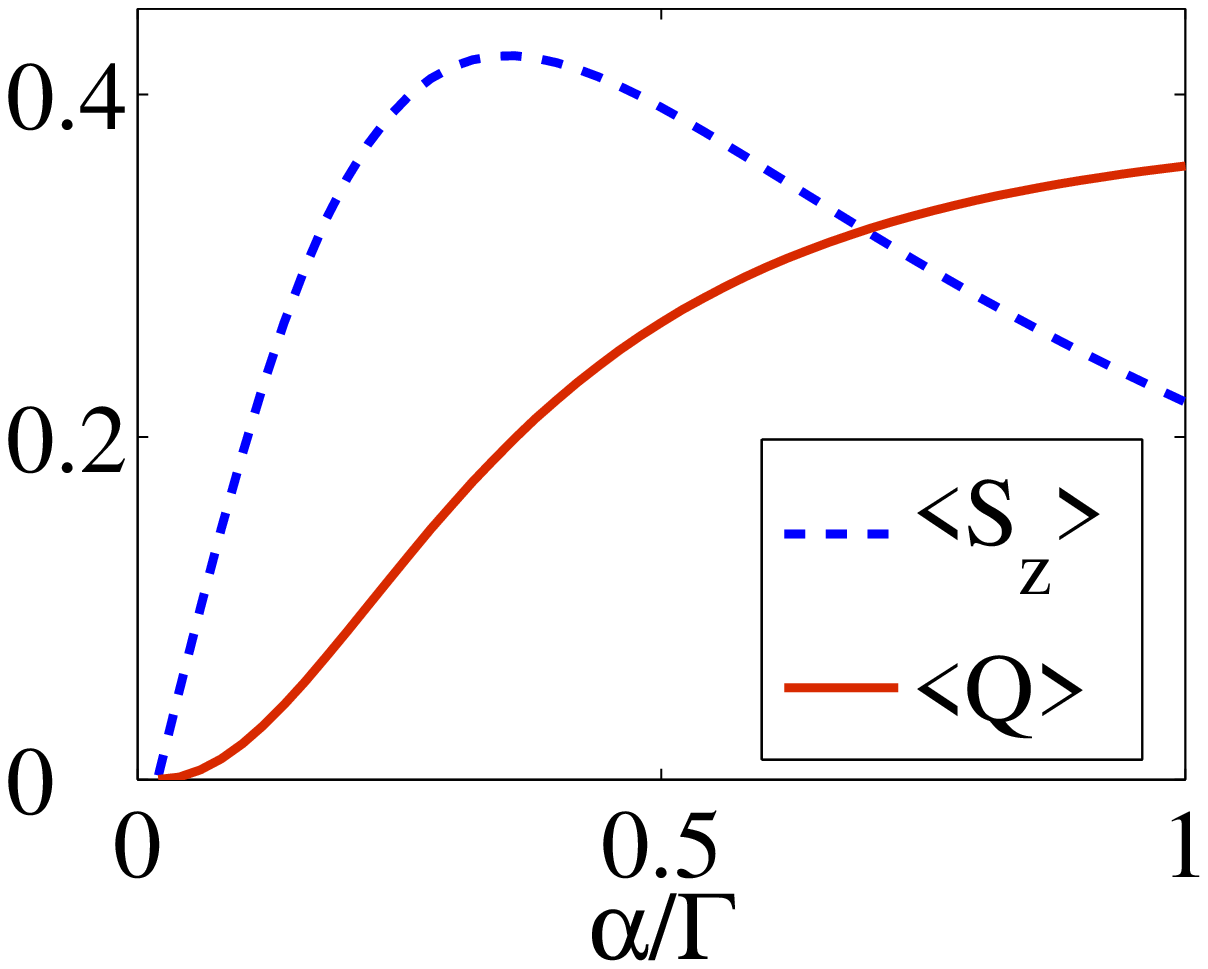}
\end{minipage}
\caption{Upper Panels: Brouwer's charge and spin field, $B^{\ve_1,\ve_2}_s$, in the $\lf((\ve_1-\ve_2),(\ve_1+\ve_2)\rg)$-plane 
for $v_{2L}=-v_{1R}=2v_{2R}=1.25 v_{1L}=\sqrt{\Gamma/(2\pi\r)}$, $\a/\Gamma=0.25$.  Lower Panels: pumped spin through a symmetric dot (left) and pumped spin and charge through a non-symmetric dot (right)  as a function 
of the SO coupling $\a$ for a triangular cycle: 
$(\ve_1+\ve_2) \in[0,10\a]$ and $(\ve_1-\ve_2) \in\lf[-(\ve_1+\ve_2),\ve_1+\ve_2\rg]$. In the symmetric dot case the tunnel couplings are  $v_{2L}=-v_{1R}=v_{2R}=v_{1L}=\sqrt{\Gamma/(2\pi\r)}$ while in the 
non-symmetric dot case they are the same as those of the upper panel.} \label{pm}
\end{figure}
For orbital energies cycles  the relevant components of the Brouwer's fields are $B_s^{\ve_1,\ve_2}$ and $B_c^{\ve_1,\ve_2}$.
 For weak SO coupling these components have a dipolar structure in the plane $\lf(\ve_1,\ve_2\rg)$, as shown in the upper panels of Fig.~\ref{pm}. As for the single-level dot, studied in Ref.~\cite{levinson},  the maxima of the charge field correspond to points of resonant transmission. 
The resonances of the spin field are instead located along the line $\ve_1^r+\ve_2^r=0$   where the renormalized energies, $\ve_1^r$ and $\ve_2^r$, are defined as
$ \ve_1^r=\ve_1\, v_{2L}v_{2R}/w$ and $\ve_2^r=\ve_2\, v_{1L}v_{1R}/w$.
As we will show, the structure of the fields and the amount of  pumped charge and spin depend very  sensitively on the  ratio between  tunneling amplitudes  and SO coupling.

To start with, let us consider the special case of a  
{\em symmetric dot} with a symmetric and an anti-symmetric orbital. 
In this case the Hamiltonian 
commutes with the operator $\pi\equiv \Sigma\otimes P$,
where  $\Sigma$ and $P$ denote the spin inversion and parity operators, and 
the tunneling amplitudes satisfy the following relations:
 $v_{1L}=v_{1R}=v_1$ and $v_{2L}=-v_{2R}=v_2$.
In this  symmetrical situation, a variation of the level energies, $\ve_1$ and $\ve_2$, 
generates a \emph{pure spin current} independently of the details of the cycle. 
In fact,  as already noted by Aleiner \emph{et al.}~\cite{aleiner00},  
parity imposes additional constraints on the scattering 
matrix that along with time-reversal symmetry lead to the vanishing of Brouwer's charge field. 
 In Fig.~\ref{pm} (lower left panel) we plot the total spin pumped through a symmetric dot with $v_1=v_2=\sqrt{\Gamma/(2\pi\r)}$ for  a triangular pumping  cycle such that $(\ve_1+\ve_2) \in[\ve_1-\ve_2,\infty]$ and $(\ve_1-\ve_2) 
\in\lf[-\infty,\infty\rg]$. As one can see the pumped spin depends non-monotonously on the ratio $\a/\Gamma$, the maximum occurring at $\alpha/\Gamma=1$. 
Finally, in Fig.~\ref{pm} (lower right panel) we plot the spin and charge pumped through a 
 non-symmetric dot in a triangular cycle bounded by the lines, $|\ve_1-\ve_2|\leq|\ve_1+\ve_2|$ and $0\leq |\ve_1+\ve_2|\leq 10 \Gamma$. 
In this case both the pumped charge and spin are non-vanishing and depend non-monotonously on the SO-coupling. 
\begin{figure}[b]
\begin{minipage}[htb]{4cm}
\centering
\includegraphics[width=3.8cm]{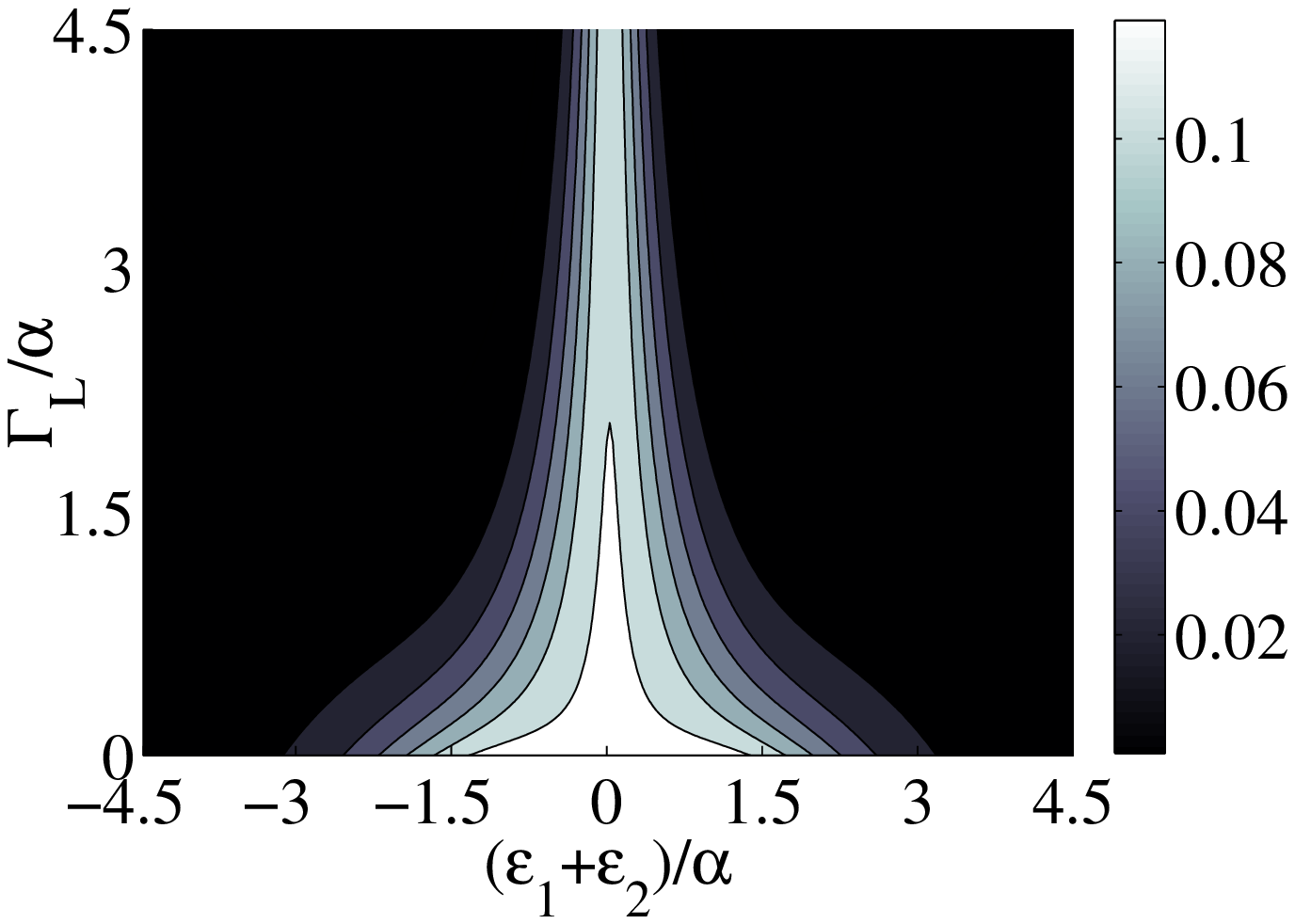}
\end{minipage}
 \hspace{3mm} 
\begin{minipage}[htb]{4cm}
\centering
\includegraphics[width=3.8cm]{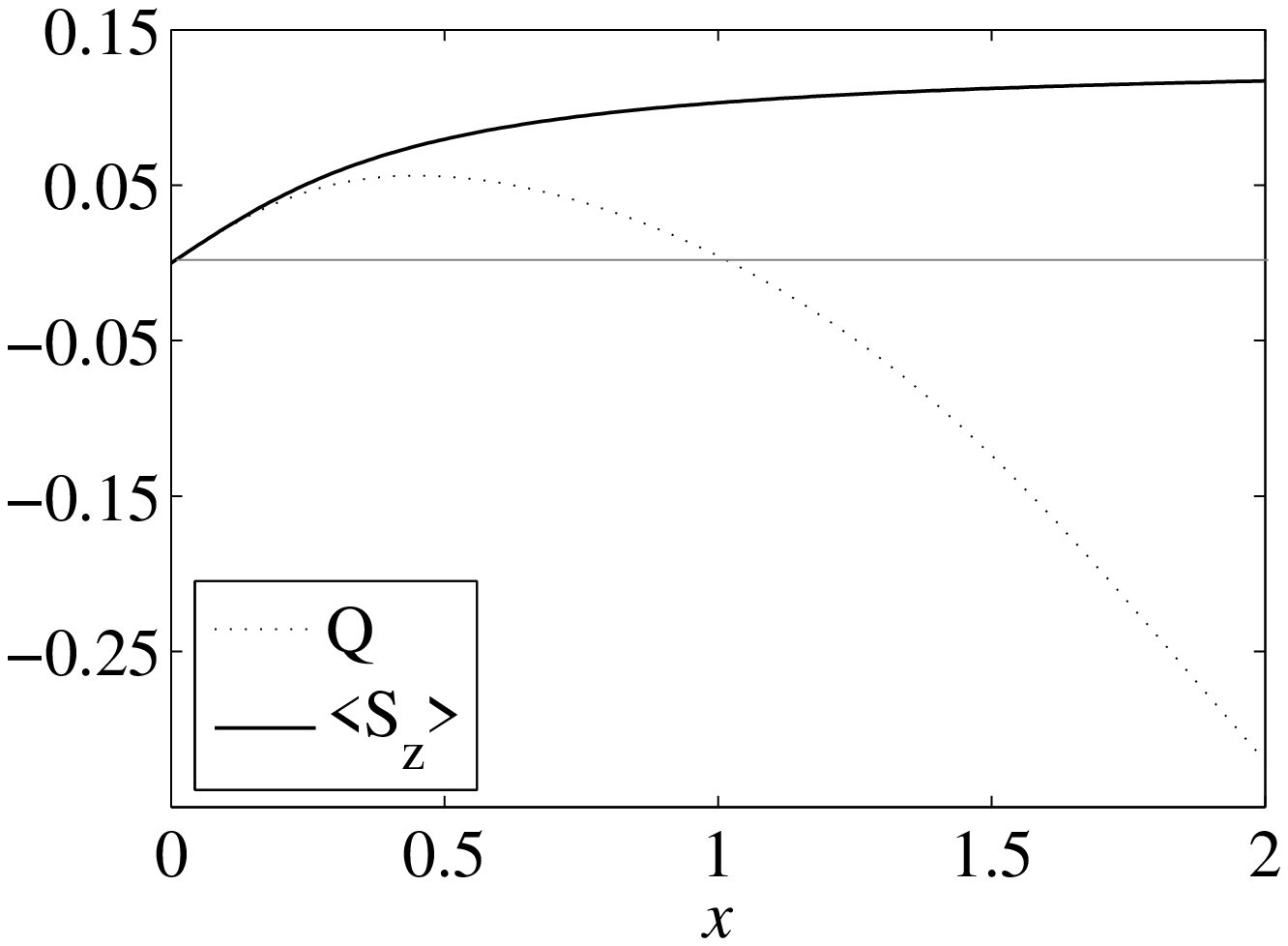}
\end{minipage}
\caption{Left Panel:  Brouwer's spin field $B^{\Gamma_L,(\ve_1+\ve_2)}_s$ in the 
$\lf(\Gamma_L,(\ve_1+\ve_2)\rg)$-plane for 
$\Gamma_{kM}=c_{kM}^2\Gamma_M$  with $k = 1,2$ and $M = L,R$, 
 $c_{1L}=c_{2L}=0.8$, $c_{1R}=0.7$, and $c_{2R}=0.6$. The two 
levels are assumed degenerate, $\ve_1-\ve_2=0$, with zero 
hybridization, $\Delta=0$, and the coupling to the right lead is
$\Gamma_R/\a=3$. Right Panel: Pumped spin and charge for a 
rectangular cycle $\Gamma_L/\a \in[0,5]$ and 
$(\ve_1+\ve_2)/\a\in[0,x]$ as a function of $x$.}\label{Lp}
\end{figure}

{\em Tunnel coupling cycles.} In the 
experiments it may be difficult to vary independently the  
energies of the two levels. For this reason we now 
consider pumping cycles where only the tunneling rates to the left 
lead and the offset of the two-levels with respect to the Fermi 
energy in the leads are varied. Specifically, we express the 
tunneling rates as follows,
 $v_{kM}=c_{kM}\sqrt{\Gamma_M/(2 \pi \rho)}$ with $k \in[1,2]$ and $M \in[L,R]$ 
and we assume that only $\Gamma_L$ can be varied using some external 
gate. The factors $c_{kM}$, describing the overlap between the $k$-th 
dot's level and the scattering states in lead $M$, are assumed to be constant during the cycle. In Fig. 
\ref{Lp} (left panel)  we plot  Brouwer's 
spin field  $B_s^{\Gamma_L,\ve_1+\ve_2}$. In this case, again, Brouwer's  charge field is non-vanishing, however, with an 
appropriate tuning of the cycle shape, we can have pure spin 
currents. This is shown in Fig. \ref{Lp} (right panel) where we plot the 
total charge and spin pumped for a rectangular cycle, $\Gamma_L/\a 
\in[0,5]$ and $(\ve_1+\ve_2)/\a\in[0,x]$ as a function of $x$.
\begin{figure}[t]
\includegraphics[width=8cm]{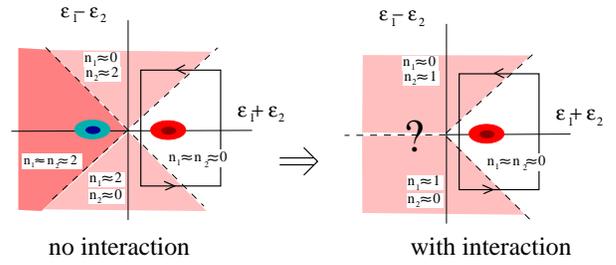}
\caption{\label{fig:phase}Sketch of the effect of interactions.}
\end{figure}

{\em Interaction effects.} To 
observe resonant spin pumping, the width of 
the two levels involved needs to be smaller or in the range of the 
level spacing. 
In this regime interaction effects, neglected so far,  become important, and 
lead to a Coulomb blockade if all levels are far away from the resonances, 
and to the formation of a  mixed valence state close to resonance. However,
in all cases, the ground state of the dot is a Fermi liquid, and therefore 
 low temperature spin puming can be described in terms of a scattering matrix, 
just as for the non-interacting system. The only difference is that the 
parameters of this scattering matrix depend in a very non-trivial way on 
the  bare model parameters. Determining the precise functional 
dependence of the scattering matrix on the gate voltages and the level 
positions is a cumbersome task beyond the scope of the 
present paper. The qualitative structure of 
the ``phase diagram''  and the spin pumping field is sketched 
in  Fig.~\ref{fig:phase}  for a dot with large Coulomb interaction. 
While the Coulomb interaction suppresses 
double occupancy for $\epsilon_{1,2} < 0$, its effects are most likely not 
crucial in the regime, $\epsilon_1,\epsilon_2> 0$, where the levels are 
only partially  occupied. There non-interacting theory is expected 
to work reasonably well, and the resonant spin pumping discussed 
here should be observable.

\emph{Conclusions.}  We have studied spin  and charge pumping in a quantum dot with spin-orbit interaction coupled to two metallic leads.
Using Brouwer's scattering approach to pumping, we first analyzed the general restrictions imposed by time-reversal symmetry on the pumped current.
We then focused on the case  of  a dot with two  resonant levels and showed that  
a spin of the order of $\hbar/2 $ can be pumped in a cycle.
We analyzed  in particular two kinds of cycles involving the offset of the levels to the Fermi energy in the leads and, either the difference between the level energies or the tunneling to the leads.
We demonstrated that, in both cases,  with an appropriate tuning of the parameters,  a controlled amount of spin and charge current can be pumped through the device. Finally we discussed the effect of Coulomb interaction, which should introduce only inessential corrections in the regime of partial dot occupation.

\emph{Acknowledgements}-We were supported by the EU FP7 under Grant
Agreement No. 238345 (GEOMDISS) and by the
DFG-Schwerpunktprogramm 1285 "Halbleiter-Spintronik".

\end{document}